\documentclass[prodmode,acmtoms,english]{acmsmall_arxiv} 
\usepackage[T1]{fontenc}
\usepackage[latin9]{inputenc}
\usepackage{float}
\usepackage{graphicx}
\usepackage{hyperref}

\usepackage{babel}
\usepackage{verbatim}
\usepackage{amsmath}
\usepackage{amsfonts}
\usepackage{amssymb}
\usepackage{ae}
\usepackage{epsfig}
\usepackage{ifthen}
\usepackage{latexsym}
\usepackage{color}
\usepackage{pbox}
\usepackage{subfig}

\usepackage{todonotes}
\usepackage{multirow}

\usepackage[ruled,linesnumbered]{algorithm2e}

\SetAlFnt{\small}
\SetAlCapFnt{\small}
\SetAlCapNameFnt{\small}
\SetAlCapHSkip{0pt}
\IncMargin{-\parindent}

\def\T{\textsuperscript{T}\ }
\def\TN{\textsuperscript{T}}



\title{SYM-ILDL: Incomplete LDL\T Factorization of Symmetric Indefinite and Skew-Symmetric Matrices}
\author{Chen Greif, Shiwen He, and Paul Liu \affil{University of British Columbia, Vancouver, Canada}}

\begin{document}

\begin{abstract}
SYM-ILDL is a numerical software package that computes incomplete LDL\T (or `ILDL') factorizations of symmetric indefinite and real skew-symmetric matrices. The core of the algorithm  is a Crout variant of incomplete LU (ILU), originally introduced and implemented for symmetric matrices by [Li and Saad, {\em Crout versions of ILU factorization with pivoting for sparse symmetric matrices}, Transactions on Numerical Analysis 20, pp. 75--85, 2005]. Our code is economical in terms of storage and it deals with real skew-symmetric matrices as well, in addition to symmetric ones. The package is written in C++ and it is templated, open source, and includes a {\sc {Matlab\texttrademark}} interface. The code includes built-in RCM and AMD reordering, two equilibration strategies, threshold Bunch-Kaufman pivoting and rook pivoting, as well as a wrapper to MC64, a popular matching based equilibration and reordering algorithm. We also include two built-in iterative solvers: SQMR, preconditioned with ILDL, and MINRES, preconditioned with a symmetric positive definite preconditioner based on the ILDL factorization.
\end{abstract}

\maketitle

\section{Introduction}

For the numerical solution of symmetric and real skew-symmetric linear systems of the form $$Ax = b,$$ stable (skew-)symmetry-preserving decompositions of $A$ often have the form $$P A P^T = LDL^{T},$$ where $L$ is a (possibly dense) lower triangular matrix and $D$ is a block-diagonal matrix with $1$-by-$1$ and $2$-by-$2$ blocks \cite{B,BK}. The matrix $P$ is a permutation matrix, satisfying $P P^T =I$,  and the right-hand side vector $b$ is permuted accordingly: in practice we solve $(P A P^T) (P x) = P b.$

In the context of incomplete LDL\T (ILDL) decompositions of sparse and large matrices for preconditioned iterative solvers, various element-dropping strategies are commonly used to impose sparsity of the factor, $L$. Fill-reducing reordering strategies are also used to encourage the sparsity of $L$, and various scaling methods are applied to improve conditioning. For a symmetric linear system, several methods have been developed. Approaches have been proposed which perturb or partition $A$ so that incomplete Cholesky may be used \cite{LM,ORBAN,SCOTT}. While \cite{LM} is designed for positive definite matrices, the recent papers of \cite{ORBAN} and \cite{SCOTT} are applicable to a large set of $2 \times 2$ block structured indefinite systems.

We present SYM-ILDL --- a software package based on a left-looking Crout version of LU,  which is stabilized by pivoting strategies such as Bunch-Kaufman and rook pivoting. The algorithmic principles underlying our software are based on (and extend) an incomplete LDL\T factorization approach proposed by \cite{ls2005}, which itself extends work by \cite{lsc2005} and \cite{js}.
We offer the following new contributions:
\begin{itemize}
	\item A Crout-based incomplete LDL\T factorization for real skew-symmetric matrices is introduced in this paper for the first time. It features a similar mechanism to the one for symmetric indefinite matrices, but there are notable differences. Most importantly, for real skew-symmetric matrices the diagonal elements of $D$ are zero and the pivots are always $2 \times 2$ blocks.

     \item We offer two integrated preconditioned solvers. The first solver is a preconditioned MINRES solver, specialized to our ILDL code. The main challenge here is to design a positive definite preconditioner, even though ILDL produces an indefinite (or skew-symmetric) factorization. To that end, for the symmetric case we implement the technique presented in \cite{Gill}. For the skew-symmetric case we introduce a positive definite preconditioner based on exploiting the simple $2 \times 2$ structure of the pivots. The second solver is a preconditioned SQMR solver, based on the work of \cite{sqmr}. For SQMR, we use ILDL to precondition it directly.

    \item The code is written in C++, and is templated and easily extensible. As such, it can be easily modified to work in other fields of numbers, such as $\mathbb{C}$. SYM-ILDL is self-contained and it includes implementations of reordering methods (AMD and RCM), equilibration methods (in the max-norm, 1-norm, and 2-norm), and pivoting methods (Bunch-Kaufman and rook pivoting). Additionally, we provide a wrapper that allows the user to use the popular {\tt HSL\_MC64} library to reorder and equilibrate the matrix. To facilitate ease of use, a {\sc {Matlab\texttrademark}} MEX file is provided which offers the same performance as the C++ version. The MEX file simply bundles the C++ library with the {\sc {Matlab\texttrademark}} interface, that is, the main computations are still done in C++.
\end{itemize}

Incomplete factorizations of symmetric indefinite matrices have received much attention recently and a few numerical packages have been developed in the past few years.
\cite{SCOTT} have developed a numerical software package based on signed incomplete Cholesky factorization preconditioners due to \cite{LM}.
For saddle-point systems, \cite{SCOTT} have extended their
limited memory incomplete Cholesky algorithm~\cite{sctu:2014b} to a signed incomplete Cholesky
factorization. Their approach builds on the ideas of \cite{TIS} and \cite{KAP}.
In the case of breakdown (a zero pivot), a global shift is applied (see also \cite{LM}).

\citeN[Section 6.4]{SCOTT} have made comparisons with our code, and have found that in general, the two codes are comparable in performance for several of the test problems, whereas for some of the problems each code outperforms the other. However, the package they compared was an earlier release of SYM-ILDL. Given the numerous improvements made upon the code since then, we repeat their comprehensive comparisons and show that SYM-ILDL now performs better than the preconditioner of \cite{SCOTT}.

\cite{ORBAN} has developed {\tt LLDL}, a generalization of the limited-memory Cholesky factorization of \cite{LM} to the symmetric indefinite case with special interest in symmetric quasi-definite matrices. The code generates a factorization of the form $\mathrm{L D L}^\mathrm{T}$ with $D$ diagonal. We are currently engaged in a comparison of our code to {\tt LLDL}.

The remainder of this paper is structured as follows. In Section \ref{sec:crout} we outline a Crout-based factorization for symmetric and skew-symmetric matrices, symmetry-preserving pivoting strategies, equilibration approaches and reordering strategies. In Section~\ref{sec:MINRES} we discuss how to modify the output of SYM-ILDL to produce a positive definite preconditioner for MINRES. In Section~\ref{sec:impl} we discuss the implementation of SYM-ILDL, and how the pivoting strategies of Section \ref{sec:crout} may be efficiently implemented within SYM-ILDL's data structures. Finally, we compare SYM-ILDL with other software packages and show the performance of SYM-ILDL on some general (skew-)symmetric matrices and some saddle-point matrices in Section~\ref{sec:results}.

\section{LDL and ILDL factorizations}
\label{sec:crout}

SYM-ILDL uses a Crout variant of LU factorization. To maintain stability, SYM-ILDL allows the user to choose one of two symmetry-preserving pivoting strategies: Bunch-Kaufman partial pivoting \cite{BK} (Bunch in the skew-symmetric case \cite{B}) and rook pivoting. The details of the factorization and pivoting procedures, as well as simplifications for the skew-symmetric case, are provided in the following sections. See also  \cite{Duff2009} for more details on the use of direct solvers for solving skew-symmetric matrices.

\subsection{Crout-based factorizations}
\label{sec:crout-fact}

The Crout order is an attractive way for computing an ILDL factorization of symmetric or skew-symmetric matrices, because it naturally preserves structural symmetry, especially when dropping rules for the incomplete factorization are applied. As opposed to the IKJ-based approach \cite{ls2005}, Crout relies on computing and applying dropping rules to a column of $L$ and a row of $U$ simultaneously. The Crout procedure for a symmetric matrix is outlined in Algorithm~\ref{alg:ldlcrout}, using a delayed update procedure for the factors which is laid out in Algorithm~\ref{alg:ldlupdate}. (As shown in Algorithm~\ref{alg:ldlcrout}, the procedure in Algorithm~\ref{alg:ldlupdate} may be called multiple times when various pivoting procedures are employed.)

\begin{algorithm}[h]
  \caption{$k$-th column update procedure}
  \label{alg:ldlupdate}
\SetAlgoNoLine
  \KwIn{A symmetric matrix $A$, partial factors $L$ and $D$, matrix size $n$, current column index $k$ }
  \KwOut{Updated factors $L$ and $D$}
      $L_{k:n,k} \leftarrow A_{k:n,k}$ \\
      $i \leftarrow 1$ \\
      \While{$i < k$} {
        $s_i \leftarrow$  size of the diagonal block with $D_{i,i}$ as its top left corner \\
        $L_{k:n,k} \leftarrow L_{k:n,k}-L_{k:n,i:i+s_i-1}D_{i:i+s_i-1,i:i+s_i-1}^{-1} L_{k,i:i+s_i-1}^T$ \\
        $i \leftarrow i+s_i$ \\
  }
\end{algorithm}

\begin{algorithm}[h]
  \caption{Crout factorization, LDL\TN C}
  \label{alg:ldlcrout}
\SetAlgoNoLine
  \KwIn{A symmetric matrix $A$}
  \KwOut{Matrices $P$, $L$, and $D$, such that $PAP \approx LDL^T$}
    $k\leftarrow1$ \\
    $L \leftarrow \mathbf{0}$ \\
    $D \leftarrow \mathbf{0}$ \\
    \While{$k<n$}{
      Call Algorithm \ref{alg:ldlupdate} to update $L$ \\
      Find a pivoting matrix in $A_{k:n,k:n}$ and permute $A$ and $L$ accordingly \\
      $s \leftarrow $ size of the pivoting matrix \\
      \If{$s=2$}{
		Update the $k+1$-th column of $L$
      }
      $D_{k:k+s-1,k:k+s-1} \leftarrow L_{k:k+s-1, k:k+s-1}$ \\
      $L_{k:n,k:k+s-1} \leftarrow L_{k:n,k:k+s-1}D_{k:k+s-1,k:k+s-1}^{-1}$ \\
      Apply dropping rules to $L_{k+s:n,k:k+s-1}$ \\
      $k \leftarrow k+s$ \\
  }
\end{algorithm}

For computing the ILDL factorization, we apply dropping rules; see line 10 of Algorithm~\ref{alg:ldlcrout}. These are the standard rules: we drop all entries below a pre-specified tolerance (referred to as {\tt drop\_tol} throughout the paper), multiplied by the norm of a column of $L$, keeping up to a pre-specified maximum number of the largest nonzero entries in every column. We use here the term {\tt fill\_factor} to signify the maximum allowed ratio between the number of nonzeros in any column of $L$ and the average number of nonzeros per column of $A$.

In Algorithm~\ref{alg:ldlcrout}, the $s \times s$ pivot is typically $1 \times 1$ or $2 \times 2$, as per the strategy devised by \cite{BK}, which we briefly describe next.

\subsection{Symmetric partial pivoting}

Pivoting in the symmetric or skew-symmetric setting is challenging, since we seek to preserve the (skew-)symmetry and it is not sufficient to use $1 \times 1$ pivots to maintain stability.
Much work has been done in this front; see, for example, \cite{DuffErismanReid1989,dgrsr1991,HoggScott2014} and the references therein.

\cite{BK} proposed a partial pivoting strategy for symmetric matrices, which relies on finding $1 \times 1$ and $2 \times 2$ pivots. The cost of finding a pivot is $\mathcal{O}(n)$, as it only involves searching up to two columns. We provide this procedure in Algorithm~\ref{alg:ldlsymmetricpp}. For all pivoting algorithms below, we assume that we are pivoting on the Schur complement (i.e. column 1 is the $k$-th column if we are on the $k$-th step of Algorithm~\ref{alg:ldlcrout}).

\begin{algorithm}[ht]
  \caption{Bunch-Kaufman LDL\T using partial pivoting strategy}
  \label{alg:ldlsymmetricpp}
\SetAlgoNoLine
  $\alpha \leftarrow (1+\sqrt{17})/8 \hspace{2pt} (\approx 0.64)$ \\
  $\omega_1 \leftarrow $ maximum magnitude of any subdiagonal entry in column 1 \\
    \eIf{$|a_{11}| \ge \alpha \omega_1$} {
       Use $a_{11}$ as a $1 \times 1$ pivot $(s = 1)$ 
    }  {
         $r \leftarrow $ row index of first (subdiagonal) entry of maximum magnitude in column 1 \\
         $\omega_r \leftarrow $ maximum magnitude of any off-diagonal entry in column $r$ \\
        \uIf{$|a_{11}| \omega_r \ge \alpha \omega_1^2$} {
          Use $a_{11}$ as a $1 \times 1$ pivot ($s = 1)$  \\ 
         } \uElseIf {$|a_{rr}| \ge \alpha \omega_r$} {
         	 	Use $a_{rr}$ as a $1 \times 1$ pivot ($s = 1$, swap rows and columns 1, $r$) 
         } \Else {
          Use
	$\begin{pmatrix}
	a_{11} & a_{r1} \\
	a_{r1} & a_{rr} \\
	\end{pmatrix}$
	as a $2 \times 2$ pivot $(s = 2$, swap rows and columns 2, $r$)  
      }
   }
\end{algorithm}

The  constant $\alpha = (1+\sqrt{17})/8$  in line 1 of the algorithm controls the growth factor, and $a_{ij}$ is the $ij$-th entry of the matrix $A$ after computing all the delayed updates in Algorithm~\ref{alg:ldlupdate} on column $i$.
Although the partial pivoting strategy is backward stable \cite{Higham}, the possibly large elements in the unit lower triangular matrix $L$ may cause numerical difficulty. Rook pivoting provides an alternative that in practice proves to be more stable, at a modest additional cost. This procedure is presented in Algorithm~\ref{alg:ldlsymmetricrp}. The algorithm searches the pivots of the matrix in spiral order until it finds an element that is largest in absolute value in both its row and its column, or terminates if it finds a relatively large diagonal element. Although theoretically rook pivoting could traverse many columns, we have found that it is as fast as Bunch-Kaufman in practice, and we use it as the default pivoting scheme of SYM-ILDL.

\begin{algorithm}[ht]
  \caption{LDL\T using rook pivoting strategy}
  \label{alg:ldlsymmetricrp}
 \SetAlgoNoLine
  $\alpha \leftarrow (1+\sqrt{17})/8 \hspace{2pt} (\approx 0.64)$ \\
  $\omega_1 \leftarrow $ maximum magnitude of any subdiagonal entry in column 1 \\
    \eIf{$|a_{11}| \ge \alpha \omega_1$} {
       Use $a_{11}$ as a $1 \times 1$ pivot $(s = 1)$ 
    }  {
      $i \leftarrow 1$ \\
      \While{a pivot is not yet chosen}{
        $r \leftarrow $ row index of first (subdiagonal) entry of maximum magnitude in column $i$ \\
        $\omega_r \leftarrow $ maximum magnitude of any off-diagonal entry in column $r$ \\
        \uIf{$|a_{rr}| \ge \alpha \omega_r$} {
          Use $a_{rr}$ as a $1 \times 1$ pivot ($s = 1$, swap rows and columns 1 and $r$)
        } \uElseIf {$\omega_i = \omega_r$} {
            Use
	$\begin{pmatrix}
	a_{ii} & a_{ri} \\
	a_{ri} & a_{rr} \\
	\end{pmatrix}$
	as a $2 \times 2$ pivot ($s = 2$, swap rows and columns 1 and $i$, and 2 and $r$)
        } \Else {
           $i \leftarrow r$ \\
           $\omega_i \leftarrow\omega_r$
        }
    }
  }
\end{algorithm}

\subsection{Equilibration and reordering strategies}
\label{sec:equib}

In many cases of practical interest, the input matrix is ill-conditioned. For these cases, equilibration schemes have been shown to be effective in lowering the condition number of the matrix. Symmetric equilibration schemes rescale entries of the matrix by computing a diagonal matrix $D$ such that $D A D$ has equal row norms and column norms.

SYM-ILDL offers three equilibration schemes. Two of the equilibration schemes are built-in: Bunch's equilibration in the max norm \cite{BunchEquil} and Ruiz's iterative equilibration in any $L_p$-norm \cite{RuizEquil}. Additionally, a wrapper is provided so that one can use MC64, a matching-based reordering and equilibration algorithm.

\subsubsection{Bunch's equilibration}
Bunch's equilibration allows the user to scale the max norm of every row and column to 1 before factorization. Let $T$ be the lower triangular part of $A$ in absolute value (diagonal included), that is, $T_{ij}=|A_{ij}|, \ 1 \leq j \leq i \leq n$. Then Bunch's algorithm runs in $\mathcal{O}({\rm nnz}(A))$ time, and is based on the following greedy procedure: \\
For $1\leq i\leq n$, set
\[
D_{ii}:=\left(\max\left\{ \sqrt{T_{ii}},\max_{1\leq j\leq i-1}D_{jj}T_{ij}\right\} \right)^{-1}.
\]

\subsubsection{Ruiz's equilibration}
Ruiz's equilibration allows the user to scale every row and column of the matrix to 1 in any $L_p$ norm, provided that $p\geq 1$ and the matrix has support \cite{RuizEquil}. For the max norm, Ruiz's algorithm scales each column's norm to within $\varepsilon$ of 1 in $\mathcal{O}({\rm nnz}(A) \log \frac{1}{\varepsilon})$ time for any given tolerance $\varepsilon$. We use a variant of Ruiz's algorithm that is similar in spirit, but produces different scaling factors.

Let $r(A,i)$ and $c(A,i)$ denote the $i$ -th row and column of $A$ respectively, and let $D(i,\alpha)$ to be the diagonal matrix
with $D_{jj}=1$ for all $j\neq i$ and $D_{ii}=\alpha$. Using this notation, our variant of Ruiz's algorithm is shown in Algorithm \ref{algo:alg3}.

\begin{algorithm}[h]

\caption{Equilibrating general matrices in the max-norm \label{algo:alg3}}
\SetAlgoNoLine
\KwIn{A general matrix $A$}
\KwOut{Diagonal matrices $R$ and $C$ such that $RAC$ has max-norm 1 in every row and column}

$R\leftarrow \bf{I}$ \\
$C \leftarrow \bf{I}$ \\
$\tilde{A} \leftarrow A$ \\
\While { $R$ {\bf and} $C$ have not yet converged } {
	\For { $i$ := 1 to $n$ } {
		$\alpha_r \leftarrow \frac{1}{\sqrt{||r(\tilde{A},i)||_\infty}}$ \\
		$\alpha_c \leftarrow \frac{1}{\sqrt{||c(\tilde{A},i)||_\infty}}$ \\
		$R \leftarrow R\cdot D(i,\alpha_r)$ \\
		$C \leftarrow C\cdot D(i,\alpha_c)$ \\
		$\tilde{A} \leftarrow D(i,\alpha_r)\tilde{A}D(i,\alpha_c)$  \\
	}
}
\end{algorithm}

Our presentation differs from Ruiz's original algorithm in that it operates on one row and column at a time as opposed to operating on the entire matrix in each iteration. We implemented the algorithm this way as it naturally adapts to our storage structures; our code is more easily amenable to single column operations rather than matrix-vector products. Hence Ruiz's original implementation and ours produce quite different scaling matrices. However, a proof of correctness similar to that of of Ruiz's algorithm applies, with the same guarantee for the running time.

\subsubsection{Matching-based equilibration}
The use of weighted matchings provides an effective technique to improve the stability of computing factorizations. In many cases, reorderings based on weighted matchings provided an effective form of static pivoting for tough indefinite symmetric problems \cite{schenk}. Our code provides a wrapper to the well-known {\tt HSL\_MC64} software package, which implements a matching-based equilibration algorithm. When MC64 is installed, our code will use a symmetrized variant of it to generate a scaling matrix that scales the max norm of every row and column to 1. More details regarding the functionality of MC64 can be found in \cite{Duff01, duffpralet05} and in the manual of the HSL Mathematical Software Library.

\subsubsection{Comparison of equilibration strategies}
Ruiz's strategy seems to perform well in terms of preserving diagonal dominance when no reordering strategy is used. In fact, we have observed that for certain skew-symmetric systems, Ruiz's equilibration leads to convergence of the iterative solver, while Bunch's approach does not. On the other hand, Bunch's  equilibration strategy is faster, being a one-pass procedure. When MC64 is available, its speed and scaling are comparable with Bunch's algorithm. However there are some matrices in our test suite for which MC64 provides a suboptimal equilibration. In our experiments we use Bunch's algorithm as the default.

\subsubsection{Fill-reducing reorderings}
After equilibration, we carry out a reordering strategy. The user is given the option of choosing from Approximate Minimum Degree (AMD) \cite{AMD}, Reverse Cuthill-McKee (RCM)  \cite{RCM}, and MC64. AMD and RCM are built into SYM-ILDL, but MC64 requires the installation of an external library. Whereas RCM and AMD are meant to reduce fill, MC64 computes a symmetric reordering of the matrix so that larger elements are placed near the diagonal. This has the effect of improving stability during the factorization, but may increase fill. Though there are matrices in our tests for which MC64 reordering is effective, we have found reducing fill to be more important, as our pivoting procedures deal with stability issues already. For the purpose of improving diagonal dominance while reducing fill, a common strategy is to combine MC64 with a fill-reducing reordering such as AMD or METIS \cite{sg2006,schenk}. We use the procedure described in \cite{schenk}, which first preprocesses the matrix with MC64 and then compresses $2\times 2$ pivots identified during the matching. The rows/columns corresponding to the pivots have their zero patterns merged and is replaced by a single row/column. Then AMD is run on the condensed matrix, after which the pivots are expanded back into two rows and columns. This procedure is implemented in {\tt HSL\_MC80}. Although MC80 is not built into our package, the results obtained by MC80 are comparable to those obtained by AMD and MC64. These results can be found in the appendix. For reducing fill, we have found both MC80 and AMD to be effective for our test cases. As MC80 requires an external library, AMD is set as the default in the code.

\subsection{LDL and ILDL factorizations for skew-symmetric matrices}

The real skew-symmetric case is different than the symmetric indefinite case in the sense that here, we must always use $2 \times 2$ pivots, because diagonal elements of real skew-symmetric matrices are zero. This simplifies both the Bunch-Kaufman and the Rook pivoting procedure: we have only one case in both scenarios. Algorithm~\ref{alg:ldlskewsymmetricrp} illustrates the simplification for rook pivoting (the simplification for Bunch is similar). Furthermore, as opposed to a typical  $2 \times 2$ symmetric matrix, which is defined by three parameters, the analogous real skew-symmetric matrix is defined by one parameter only.
As a result, at the $k$th step, the computation of the multiplier and the subsequent update of pair of columns associated with the pivoting operation can be expressed as follows:
\begin{align*}
A_{k+2:n,k:k+1}A_{k:k+1,k:k+1}^{-1}&=A_{k+2:n,k:k+1}\left(\begin{array}{cc}0&-a_{k+1,k}\\a_{k+1,k}&0\end{array}\right)^{-1}\\
&=\frac{1}{a_{k+1,k}}A_{k+2:n,k:k+1}\left(\begin{array}{cc}0&1\\-1&0\end{array}\right),
\end{align*}
which can be trivially computed by swapping  columns $k$ and $k+1$ and scaling.

\begin{algorithm}[ht!]
  \caption{LDL\T using rook pivoting strategy for skew-symmetric matrices}
  \label{alg:ldlskewsymmetricrp}
\SetAlgoNoLine
    $\omega_1 \leftarrow $ maximum magnitude of any subdiagonal entry in column 1 \\
    $i \leftarrow 1$ \\
      \While{a pivot is not yet chosen}{
        $r \leftarrow $ row index of first (subdiagonal) entry of maximum magnitude in column $i$ \\
        $\omega_r \leftarrow $ maximum magnitude of any off-diagonal entry in column $r$ \\
        \eIf{$\omega_i = \omega_r$} {
         Use
$\begin{pmatrix}
0 & -a_{ri} \\
a_{ri} & 0 \\
\end{pmatrix}$
as a $2 \times 2$ pivot (swap rows and columns 1 and $i$, and 2 and $r$)
        } {
           $i \leftarrow r$ \\
           $\omega_i \leftarrow\omega_r$
        }}
\end{algorithm}

The ILDL factorization for skew-symmetric matrices can thus be carried out similarly to the manner in which it is developed for symmetric indefinite matrices, but the eventual algorithm gives rise to the above described simplifications.
Skew-symmetric matrices are often ill-conditioned, and we have experimentally found that computing a numerical solution effectively for those systems is challenging. More details are provided in Section \ref{sec:results}.

\section{Iterative solvers for SYM-ILDL}
\label{sec:MINRES}

In SYM-ILDL we implement two preconditioned iterative solvers: SQMR and MINRES. For SQMR, we can use the incomplete $LDL^{T}$ factorization as a preconditioner directly. For MINRES, we require the preconditioner to be positive definite, and modify our $LDL^{T}$ as described in Section \ref{sec:MINRES}.

In our experiments, SQMR usually took fewer iterations to converge, with the same arithmetic cost per iteration. However, we found MINRES to be useful and more stable in difficult problems where SQMR stagnates (see Section \ref{sec:results}). In these problems, MINRES returns a solution vector with a much smaller residual than SQMR in fewer iterations.

\subsection{A specialized preconditioner for MINRES}


We describe below techniques for generating MINRES preconditioned iterations, using positive definite versions of the incomplete factorization.
 For the symmetric indefinite case, we apply the method presented in \cite{Gill}. Given
$ M = L D L^T, $
let us focus our attention on the various options for the blocks of $D$. Our ultimate goal is to modify $D$ and $L$ such that $D$ is diagonal with only $1$ or $-1$ as its diagonal entries. If a block of the matrix $D$ from the original LDL factorization was $2 \times 2$, then the corresponding modified (diagonal) block would become
$$ \begin{pmatrix}
    \pm 1 & 0 \\
    0 & \mp 1
    \end{pmatrix}.$$
     For a diagonal entry of $D$ that appears as a $1 \times 1$ block, say, $d_{i,i}$, we rescale the $i$th row of $L$: $L(i,:) \to L(i,:) \sqrt{|d_{i,i}|}$. We can then set the new value of $d_{i,i}$ as $\text{sgn}(d_{i,i})$. In practice there is no need to perform a multiplication of a row of $L$ by $\sqrt{|d_{i,i}|}$; instead, this scalar is stored separately and its multiplicative effect is computed as an $\mathcal{O}(1)$ operation for every matrix vector product.

Now, consider a $2 \times 2$ block of $D$, say $D_j$. For this case, we compute the eigendecomposition $$D_j = Q_j \Lambda_j Q_j^T, $$
and similarly to the case of a $1 \times 1$ block, we {\em implicitly} rescale two rows of $L$ by $Q_j \sqrt{|\Lambda_j|}$. This means that $L$ is no longer triangular; it is in fact lower Hessenberg, since some values above the main diagonal may become nonzero. But the solve is just as straightforward, since the decomposition is explicitly given.

Since $L$ was originally a unit lower triangular matrix that was scaled by positive scalars, $L L^T$ is symmetric positive definite and we use it as a preconditioner for MINRES. Note that if we were to compute the {\em full} $L D L^T$ decomposition and scale $L$ as described above, then MINERS would converge within two iterations (in the absence of roundoff errors), thanks to the two eigenvalues of $D$, namely $1$ and $-1$.

  In the skew-symmetric case, we may use a specialized version of MINRES \cite{gv2009}.
  We only have $2 \times 2$ blocks, and for those, we know that
$$ \begin{pmatrix} 0 & a_{j,j} \\
                    -a_{j,j} & 0
    \end{pmatrix}
    =
    \begin{pmatrix}  \sqrt{|a_{j,j}|} & 0 \\
                     0 & \sqrt{|a_{j,j}|}
    \end{pmatrix}
    \begin{pmatrix} 0 & \pm 1 \\
                    \mp 1 & 0
    \end{pmatrix}
    \begin{pmatrix}  \sqrt{|a_{j,j}|} & 0 \\
                     0 & \sqrt{|a_{j,j}|}
    \end{pmatrix} .
    $$
Therefore, we do not need an eigendecomposition (as in the symmetric case), and instead we just scale the two affected rows of $L$ by $\sqrt{|a_{j,j}|} I_2$.

\begin{figure}[h]
\begin{center}
	\includegraphics[width=\textwidth]{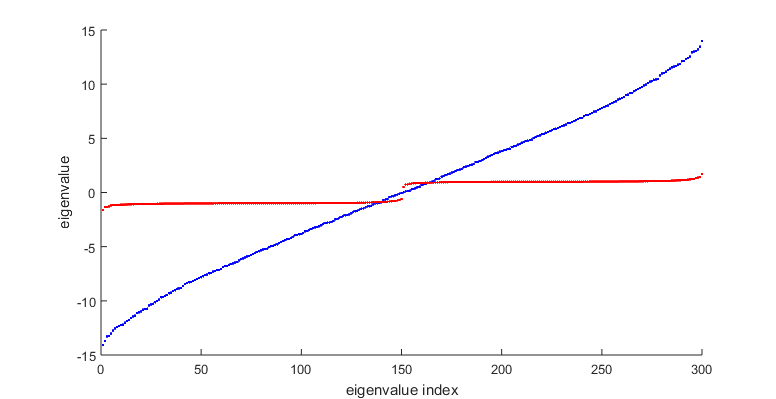}
\caption{Eigenvalues of a preconditioned symmetric random $300 \times 300$ matrix (red). Note the clustering at 1 and -1. Entries in the matrix are drawn from the standard normal distribution. The unpreconditioned eigenvalues are in blue. \label{fig:eigs}}
\end{center}
\end{figure}

Figure \ref{fig:eigs} shows the clustering effect that the proposed preconditioning approach has. We generate a random real symmetric $300 \times 300$ matrix $A$ (with entries drawn from the standard normal distribution), and compute the eigenvalues of $(LDL^T)^{-1} A$, where $L$ and $D$ are the matrices generated in the above described preconditioning procedure. Our fill factor is $2.0$ and the drop tolerance was $10^{-4}$. We note that the eigenvalue distribution in the figure is typical for other cases that were tested.

\section{Implementation}

\label{sec:impl}
\subsection{Matrix storage in SYM-ILDL}
\label{sec:matstor}

\begin{figure}[ht]
\begin{center}
	\includegraphics[width=\textwidth]{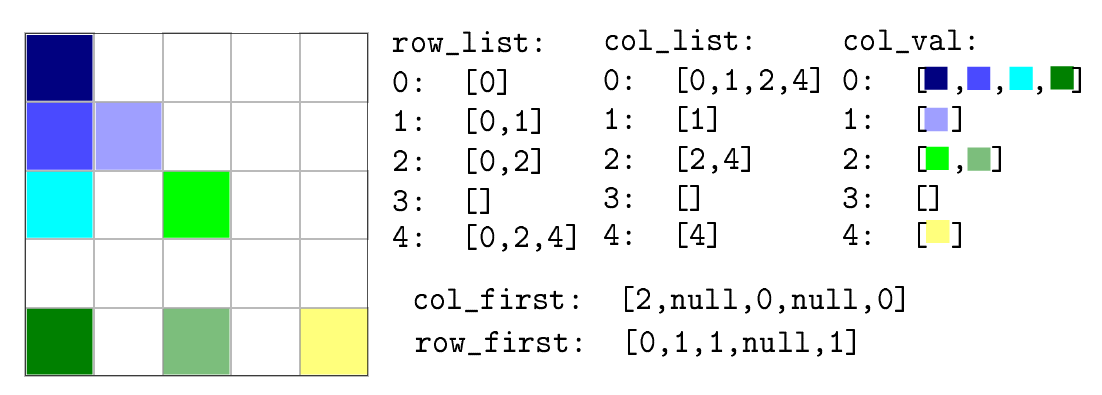}

\caption{Graphical representation of the data structures of SYM-ILDL. {\tt col\_first} and {\tt row\_first} are shown during the third iteration of the factorization. Hence {\tt col\_first} holds the values of indices in {\tt col\_list} for the first element under or on the third row of the matrix. Similarly, {\tt row\_first} holds the values of indices in {\tt row\_list} for the last element not exceeding the third column of the matrix. \label{storage}}
\end{center}

\end{figure}

Since we are dealing with symmetric or skew-symmetric matrices, one of our goals is to avoid duplicating data. At the same time, it is necessary for SYM-ILDL to have fast column access as well as fast row access. In terms of storage, we deal with these requirements by generating a format similar to standard compressed sparse column form, along with compressed sparse row form without the nonzero floating point matrix values. Matrices are stored in a {\em list-of-arrays} format. Each column is represented internally as two dynamically sized arrays, storing both its nonzero values {\tt col\_val} and row indices ({\tt col\_list}). These arrays facilitate fast random accesses as well as removals from the middle of the array (by simply swapping the value to be deleted to the end of the array, and decrementing its size by 1). Meanwhile, another array holds pairs of pointers to the two column arrays of each column. One advantage of this format is that swapping columns and deallocating their memory is much easier, as we only need to operate on the array holding column pointers.
Additionally, a row-major data structure ({\tt row\_list}) is used to maintain fast access across the nonzeros of each row (see Figure \ref{storage}).
This is obtained by representing each row internally as a single array, storing the column indices of each row in an array (the nonzero values
are already stored in the column-major representation).

Our format is an improvement over storing the full matrix in standard CSC, as used in  \cite{ls2005}. Assuming
that the row and column indices are stored in 32-bit integers and the nonzero values are stored in 64-bit doubles, this gives us an overall $33\%$ saving in storage if we were to store the factorization in-place. This is an easy modification of Algorithm~\ref{alg:ldlsymmetricpp}. In the default implementation, we find it more useful to store an equilibrated and permuted copy of the original matrix, so that we may use it for MINRES after the preconditioner is computed. An in-place version that returns only the preconditioner is included as part of our package.

\subsection{Data structures for matrix access}

In ILUC \cite{ls2005}, a bi-index data structure was developed to address two implementation difficulties in sparse matrix operations, following earlier work by \cite{Eisenstat} in the context of the Yale Sparse Matrix Package (YSMP), and \cite{js}. Our implementation uses a similar bi-index data structure, which we briefly describe below.

Internally, the column and row indices in the matrix are stored in partial order, with one array per column a
nd row. On the $k$-th iteration, elements are partially sorted so that all row indices less than $k$ are stored in a contiguous segment per column, and all row indices greater or equal to $k$ are stored in another contiguous segment. Within each segment, the elements are unsorted. This avoids the cost of sorting whenever we need to pivot.  Since elements are partially sorted, accessing specific elements
of the matrix is difficult and requires a slow linear search. Luckily,
because Algorithm~\ref{alg:ldlsymmetricpp} accesses elements in a predictable
fashion, we can speed up access to subcolumns required during the
factorization to $\mathcal{O}(1)$ amortized time. The strategy we use to speed
up matrix access is similar to that of \cite{js}. To ensure
fast access to the submatrix $L_{k+1:n,1:k}$ and the row $L_{k,:}$
during factorization, we use one additional length $n$ array: {\tt col\_first}.
On the $k$-th iteration, the $i$-th element of {\tt col\_first} holds an offset that stores the dividing index between indices less than $k$ and greater or equal to $k$. In effect, {\tt col\_first} gives us fast access to the start of the submatrix $L_{k+1:n,i}$ in {\tt col\_list}
and speeds up  Algorithm~\ref{alg:ldlupdate}, allowing
us access to the submatrix in $\mathcal{O}(1)$ time. To get fast
access to the list of columns that contribute to the update of the
$(k+1)$-st column, we use the row structure {\tt row\_list} discussed
in Section~\ref{sec:matstor}. To speed up access to {\tt row\_list}, we maintain a {\tt row\_first} array that is implemented similarly to the {\tt col\_first}. Overall, this reduces the access time of the submatrix $L_{k+1:n,1:k}$ and row $L_{k}$ down to a cost
proportional to the number of nonzeros in these submatrices.

Before the first iteration, {\tt col\_first(i)} is initialized to an array of all zeros. To ensure that {\tt col\_first(i)} stores the correct offset to the start of the
subcolumn $L_{k+1:n,i}$ on step $k$, we increment the offset for
{\tt col\_first(i)} (if needed) at the end of processing the $k$-th
column. Since the column indices in {\tt col\_list} are unsorted,
this step requires a linear search to find the smallest element in
{\tt col\_list}. Once this element is found, we swap it to the correct
spot so that the column indices for $L_{k+1:n,i}$ are in a contiguous
segment of memory. We have found it helpful to speed up the linear search by ensuring the indices of $A$ are sorted before beginning the factorization. This has the effect that $A$ remains roughly sorted when there are few pivot steps.

Similarly, we will also need to access the subrows $A_{k,1:k}$ and
$A_{r,1:k}$ during the pivoting stage (lines 11 to 15 in Algorithm~\ref{alg:ldlsymmetricpp} and Algorithm~\ref{alg:ldlsymmetricrp}). This is sped up by an analogous {\tt row\_first(i)} structure
that stores offsets to the end of the subrow $A_{i,1:k}$ ($A_{i,1:k}$ is
the memory region that encompases everything from the start of memory
for that row to {\tt row\_first(i)}). At the end of step $k$, we
also increment the offsets for {\tt row\_first} if needed.

A summary of data structures  can be found
in Table \ref{tab:datastructures}.

\renewcommand{\arraystretch}{2.5}
\begin{table}[ht]

\tbl{Variable names with data structure types \label{tab:datastructures}}{
\begin{tabular}{|c|c|c|}
\hline
\textbf{Variable name} & \textbf{Data structure type} & \textbf{Purpose}\tabularnewline
\hline
{\tt col\_first} & \emph{n} length array & Speeds up access to $L_{k+1:n,i}$, i.e., {\tt row\_list}\tabularnewline
\hline
{\tt row\_first} & \emph{n} length array & Speeds up access to $A_{i,1:k}$, i.e., {\tt col\_list}\tabularnewline
\hline
{\tt row\_list} & \emph{n} dynamic arrays (row-major) & Stores indices of $A$ across the rows\tabularnewline
\hline
{\tt col\_list} & \emph{n} dynamic arrays (col-major) & Stores indices of $A$ across the columns\tabularnewline
\hline
{\tt col\_val} & \emph{n} dynamic arrays (col-major) & Stores nonzero coefficients of $A$\tabularnewline
\hline
\end{tabular}}

\end{table}

\renewcommand{\arraystretch}{1}

\section{Numerical experiments}
\label{sec:results}

All our experiments were run single threaded, on a machine with a 2.1 GHz Intel Xeon CPU and 128 GB of RAM. In the experiments below, we follow the conventions of \cite{ls2005,lsc2005} and define the {\em fill} of a factorization as $\mathrm{nnz}(L+D+L^{T})/\mathrm{nnz}(A)$. Recall that the {\tt fill\_factor} is the maximum allowed ratio between the number of nonzeros in any column of $L$ and the average number of nonzeros per column of $A$. Therefore, the fill of our preconditioner is bounded by approximately $2\cdot \mathtt{fill\_factor}$; the factor of 2 arises from the symmetry.

\subsection{Results for symmetric matrices}

\subsubsection{Tests on general symmetric indefinite matrices}
For testing our code, we use the University of Florida (UF) collection \cite{DavisYifan2011}, as well as our own matrices. The UF collection provides a variety of symmetric matrices, which we specify in Tables \ref{tab:small} and~\ref{tab:scott}. We have used some of the same matrices that have been used in the papers \cite{ls2005,lsc2005,SCOTT}.

In Table \ref{tab:small} we show the results of experiments with  a set of matrices from~\cite{DavisYifan2011} as well as comparisons with {\sc MATLAB}'s ILUTP. The matrix dimensions go up to approximately four million, with number of nonzeros going up to approximately 100 million.
 We show timings for constructing the ILDL factorization and an iterative solution, applying preconditioned SQMR for SYM-ILDL and GMRES(20) for ILUTP with drop tolerance $10^{-3}$ for a maximum of 1000 iterations. We apply either Bunch's equilibration or MC64 scaling and either AMD or MC64 reordering (MC64R) before generating the ILDL factorization. Preconditioned SQMR is run with SYM-ILDL for a maximum of 1000 iterations. We show the best results in Table \ref{tab:small} out of the 4 possible reordering and equilibration combinations for both ILUTP and ILDL. We have also tested the ILDL preconditioner with {\tt HSL\_MC80} reordering and equilibration and have found it to be comparable with the best of the 4 combinations above. The full test data for all 4 combinations as well as tests with MC80 can be found in Table \ref{tab:all-results} of the appendix. For the incomplete factorization, we apply rook pivoting. We observe that ILDL achieves similar iteration counts with a far sparser preconditioner. Furthermore, even for cases where ILDL was beaten on iteration count, we see that the denser factor of ILUTP causes the overall solve time to be much slower than ILDL. When ILDL and ILUTP have similar fill, ILDL converges in fewer iterations. 
 
\begin{figure}[ht]
\centering
\includegraphics[width=\textwidth]{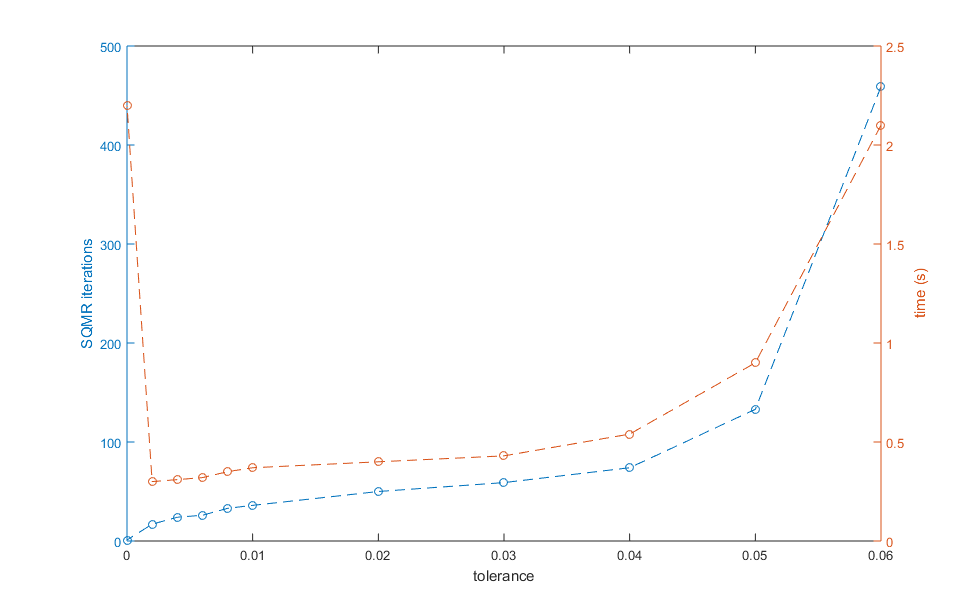}

\caption{The number of SQMR iterations and total precondition+solve time as a function of tolerance. Tests were performed on the {\tt tuma1} matrix with Bunch equilibration, AMD reordering, and {\tt fill\_factor} set to $\infty$ (to measure only the tolerance dropping rule). \label{fig:tol-v-iters}}
\end{figure}

 In Figure~\ref{fig:tol-v-iters} we examine the sensitivity of ILDL to input tolerance. We plot the number of iterations and the timings for a changing value of the tolerance. We observe the expected behavior. As the tolerance decreases, there is a tradeoff between preconditioning time and iteration count. Thus, the total computational time is high at both extremes. That said, there is a large range of values of tolerance for which both time and iterations are modest. Altogether, ILDL works well for all test cases with fairly generic parameters.

\renewcommand{\arraystretch}{1.2}
\begin{table}[ht]

\tbl{Factorization timings and SQMR iterations for test matrices \label{tab:small}}{
\begin{tabular}{lll|llll}
\hline
matrix  & $n$  & ${\rm nnz}(A)$  & fill  &  time (s) & type & iterations\tabularnewline
\hline
aug3dcqp  & $35543$  & $128115$  & $1.9$ & $0.05+0.15$ & ILDL(B+AMD) & $24$\tabularnewline
  & & & 7.3 & 2.66+0.20 & ILUTP(B+AMD) & 6\tabularnewline
\hline
bloweya  & $30004$  & $150009$  & $1.0$ & $0.07+0.02$ & ILDL(MC64+MC64R) & $3$\tabularnewline
  & & & 3.2 & 7.86+0.10 & ILUTP(B+MC64R) & 3\tabularnewline
\hline
bratu3d  & $27792$  & $173796$  & $3.8$ & $0.25+0.11$ & ILDL(B+MC64R) & $18$\tabularnewline
  & & & 8.1 & 8.50+0.54 & ILUTP(B+MC64R) & 11\tabularnewline
\hline
tuma1  & $22967$  & $87760$  & $3.0$  & $0.05+0.13$ & ILDL(MC64+MC64R) & $35$\tabularnewline
  & & & 7.8 & 2.68+0.58 & ILUTP(B+AMD) & 14\tabularnewline
\hline
tuma2  & $12992$  & $49365$  &  $3.0$ & $0.03+0.09$ & ILDL(MC64+MC64R) & $28$\tabularnewline
  & & & 6.9 & 0.72+0.23 & ILUTP(B+AMD) & 13\tabularnewline
\hline
boyd1  & $93279$  & $1211231$  & $1.0$  & $0.10+0.50$ & ILDL(B+AMD)& $3$ \tabularnewline
  & & & 0.8 & 0.26+0.86 & ILUTP(B+MC64R) & 10\tabularnewline
\hline
brainpc2  & $27607$  & $179395$  & $1.0$  & $0.31+0.10$ & ILDL(MC64+MC64R) & $31$ \tabularnewline
  & & & 0.6 & 0.54+38.7 & ILUTP(B+AMD) & NC\tabularnewline
\hline
mario001  & $38434$  & $204912$  & $3.7$ & $0.13+0.56$ & ILDL(B+MC64R) & $52$ \tabularnewline
  & & & 8.0 & 2.47+0.54 & ILUTP(B+AMD) & 8\tabularnewline
\hline
qpband  & $20000$  & $45000$  & $1.1$  & $0.008+0.004$ & ILDL(B+AMD) & $1$ \tabularnewline
  & & & 1.1 & 0.008+0.021 & ILUTP(B+AMD) & 1\tabularnewline
\hline
nlpkkt80  & $1062400$  & $28192672$  & $8.0$  & $153+53$ & ILDL(B+MC64R) & 34\tabularnewline
  & & & 4.1 & 6803+2502 & ILUTP(B+AMD) & NC\tabularnewline
\hline
nlpkkt120  & $3542400$  & $95117792$  & $8.0$  & $525+334$ & ILDL(B+MC64R) & 58 \tabularnewline
  & & & - & - & ILUTP & -\tabularnewline
\hline
\end{tabular}}

\begin{tabnote}
The experiments were run with {\tt fill\_factor} = 2.0 for the smaller matrices and {\tt fill\_factor} = 4.0 for matrices larger than one million in dimension. The tolerance was
{\tt drop\_tol} = $10^{-4}$, and we used rook pivoting to maintain stability. The iteration was terminated when the norm of the relative residual went below $10^{-6}$. The time is reported as $x+y$, where $x$ is the preconditioning time and $y$ is the iterative solver time. Times labelled with `-' took over 10 hours to run, and were terminated before completion. Iteration counts labelled with NC indicates that the problem did not converge within 1000 iterations.
\end{tabnote}

\end{table}

\subsubsection{Further comparisons with {\sc Matlab}'s ILUTP}

To show the memory efficiency of our code, we consider matrices associated with the discrete Helmholtz equation,
\begin{equation}
\label{eq:helmholtz}
-\Delta u -\alpha u = f,
\end{equation}
subject to Dirichlet boundary conditions
on the unit square, discretized using a uniform mesh of size $h$. Here we choose a moderate value of $\alpha$, so that a symmetric indefinite matrix is generated. The choice of $\alpha$ may have a significant impact on the conditioning of the matrix. In particular, if $\alpha$ is an eigenvalue, then the shifted matrix is singular. In SYM-ILDL, a singular matrix will trigger static pivoting, and may add a significant computational overhead. In our numerical experiments we have stayed away from choices of $\alpha$ such that the shifted matrix is singular. Although we could have used the same matrices as Table \ref{tab:small}, additional tests using Helmholtz matrices provide a greater degree of insight as we know its spectra and can easily control the dimension and number of non-zeros.

In Table \ref{tab:comparison} we present results for the Helmholtz model problem. We compare SYM-ILDL to {\sc Matlab}'s ILUTP. For ILUTP we used a drop tolerance of $10^{-3}$ in all test cases. For ILDL, the {\tt fill\_factor} was set to $\infty$ (since ILUTP does not limit its intermediate memory by a fill factor) while the {\tt drop\_tol} parameter was then chosen to get roughly the same fill as that of ILUTP. In the context of the ILUTP preconditioner, the {\em fill} is defined as $\mathrm{nnz}(L+U)/\mathrm{nnz}(A)$.

For both ILDL and ILUTP, GMRES(100) was used as the iterative solver and the input matrix was scaled with Bunch equilibration and reordered with AMD. During the computation of the preconditioner, the in-place version of ILDL uses only about 2/3 of the memory used by ILUTP. During the GMRES solve, the ILDL preconditioner only uses about 1/2 the memory used by ILUTP. We note that ILDL could also be used with SQMR, which has a much smaller memory footprint than GMRES.

We observe that the performance of ILDL on the Helmholtz model problem is dependent on the value of $\alpha$ chosen, but that if ILDL is given the same memory resources as ILUTP, ILDL outperforms ILUTP. The memory usage of ILUTP and ILDL are measured through the {\sc MATLAB} profiler. For $\alpha=0.3/h^2$, the ILDL approach leads to lower iteration counts even when approximately 1/2 of the memory is allocated (i.e., when the same fill is allowed), whereas for $\alpha=0.7/h^2$, ILUTP outperforms ILDL when the fill is roughly the same. If we allow ILDL to have memory usage as large as ILUTP (i.e., up to roughly 3/2 the fill), we see that ILDL clearly has lower iteration counts for GMRES.

\renewcommand{\arraystretch}{1.3}
\begin{table}[ht]


\tbl{Comparison of {\sc Matlab}'s ILUTP and SYM-ILDL for Helmholtz matrices \label{tab:comparison}}{
\begin{tabular}{ccccccccc}
\hline
matrix  & $n$  & ${\rm nnz}(A)$  & ilu fill  & ilu gmres iters & ildl fill  & ildl gmres iters\tabularnewline
\hline
$\alpha = 0.3/h^2$, Extra memory for ILUTP \tabularnewline
\hline

helmholtz80  & 6400  & $31680$  & $7.7$  & $11$ & $7.6$  & $8$\tabularnewline 

helmholtz120  & 14400  & $71520$  & $10.6$  & $13$  & $10.3$  & $8$\tabularnewline 

helmholtz160  & 25600  & $127360$  & $12.3$  & $17$  & $12.3$  & $8$ \tabularnewline 

helmholtz200  & 40000  & $199200$  & $14.1$  & $24$  & $14.0$  & $11$\tabularnewline 

\hline
$\alpha = 0.7/h^2$, Extra memory for ILUTP \tabularnewline
\hline
helmholtz80  & 6400  & $31680$  & $7.4$  & $5$ & $7.5$  & $8$\tabularnewline 

helmholtz120  & 14400  & $71520$  & $13.4$  & $10$  & $14.0$  & $18$\tabularnewline 

helmholtz160  & 25600  & $127360$  & $16.4$  & $15$  & $16.7$  & $43$ \tabularnewline 

helmholtz200  & 40000  & $199200$  & $19.5$  & $20$  & $20.8$  & $86$\tabularnewline 

\hline
$\alpha = 0.7/h^2$, Equal memory for ILDL and ILUTP \tabularnewline
\hline
helmholtz80  & 6400  & $31680$  & $7.4$  & $5$ & $11.0$  & $6$\tabularnewline 

helmholtz120  & 14400  & $71520$  & $13.4$  & $10$  & $18.6$  & $6$\tabularnewline 

helmholtz160  & 25600  & $127360$  & $16.4$  & $15$  & $22.8$  & $8$ \tabularnewline 

helmholtz200  & 40000  & $199200$  & $19.5$  & $20$  & $33.0$  & $11$\tabularnewline 

\hline
\end{tabular}}

\begin{tabnote}
 The parameter $\alpha$ in Equation~\ref{eq:helmholtz} is indicated above. GMRES was terminated when the relative residual decreased below $10^{-6}$.
\end{tabnote}

\end{table}

\subsubsection{Comparisons with {\tt HSL\_MI30}}
In Table~\ref{tab:scott} we compare our code to the code of \cite{SCOTT}, implemented in the package {\tt HSL\_MI30}. This comparison was already done in \cite{SCOTT}, with an older version of SYM-ILDL. However, with recent improvements, we see that SYM-ILDL generally takes 2-6 times fewer iterations than {\tt HSL\_MI30}. The matrices we compare with are a subset of the matrices used in the original comparison. In particular, these matrices were ones for which SYM-ILDL performed the most poorly in the original comparison. The matrices were obtained from the University of Florida matrix collection \cite{DavisYifan2011}.

The parameters used here are almost the same as in the original comparison. For {\tt HSL\_MI30}, we used the built-in {\sc MATLAB} interface, set {\tt lsize} and {\tt rsize} to 30, $\alpha(1:2)$ both to 0.1, the drop tolerances $\tau_1$ and $\tau_2$ to $10^{-3}$ and $10^{-4}$, and used the built-in {\tt MC77} equilibration (which performed the best out of all possible equilibration options, including MC64). We also tried all possible reordering options for {\tt HSL\_MI30} and found that the natural ordering performed the best. For SYM-ILDL, we used a {\tt fill\_factor} of 12.0, {\tt drop\_tol} of 0.003, as in the original comparison. The only difference between the original comparison and this one is that rook pivoting is used for stability and MC80 is used for equilibration and reordering. We have also performed additional tests using MC64 for equilibration and AMD for reordering and have found comparable number of iterations with higher fill. All tests can be found in the appendix.

\begin{table}[ht]

\tbl{GMRES comparisons between SYM-ILDL and {\tt HSL\_MI30}\label{tab:scott} }{
\begin{tabular}{c|cc|ccc|ccc}
\hline
Matrix name & $n$ & ${\rm nnz}(A)$ & $\text{fill}_\text{MI30}$ & MI30 iters & time (s) & $\text{fill}_\text{SYM-ILDL}$ & SYM-ILDL iters & time (s) \tabularnewline
\hline
c-55  & 32780  & 403450  & 3.45 & 49 & 1.25+0.94 & 2.95 & 15 & 0.23+0.15\tabularnewline
c-59  & 41282  & 480536  & 3.62 & 70 & 1.59+1.84 & 2.99 & 15 & 0.36+0.20\tabularnewline
c-63  & 44234  & 434704  & 4.10 & 51 & 1.53+1.23 & 2.92 & 15 & 0.29+0.21\tabularnewline
c-68  & 64810  & 565996  & 4.12 & 37 & 1.87+1.12 & 2.31 & 9 & 0.31+0.17\tabularnewline
c-69  & 67458  & 623914  & 4.33 & 43 & 4.07+1.47 & 2.65 & 9 & 0.35+0.18\tabularnewline
c-70  & 68924  & 658986  & 4.26 & 38 & 3.77+1.30 & 2.67 & 11 & 0.40+0.24\tabularnewline
c-71  & 76638  & 859520  & 3.58 & 61 & 3.93+2.71 & 3.00 & 12 & 0.74+0.32\tabularnewline
c-72  & 84064  & 707546  & 4.18 & 54 & 3.05+2.40 & 2.69 & 9 & 0.40+0.31\tabularnewline
c-big & 345241 & 2340859 & 4.82 & 67 & 23.4+25.3 & 2.54 & 8 & 1.20+0.93\tabularnewline
\hline
\end{tabular}}

\begin{tabnote}
For each test case, we report the time it takes to compute the preconditioner, as well as the GMRES time and the number of GMRES iterations. The time is reported as $x+y$, where $x$ is the preconditioning time and $y$ is the GMRES time. GMRES was terminated when the relative residual decreased below $10^{-6}$.
\end{tabnote}

\end{table}

We note that although we set the {\tt fill\_factor} to be 12.0 in all comparisons with {\tt HSL\_MI30}, SYM-ILDL can have similar performance with a much smaller {\tt fill\_factor}.

\subsection{Results for skew-symmetric matrices}

 We test with  a skew-symmetrized version of a model convection-diffusion equation, which is a discrete version of
\begin{equation}
\label{eq:convdiff}
-\Delta u + (\sigma, \tau, \mu) \nabla u = f,
\end{equation}
with Dirichlet boundary conditions on the unit square, discretized using a uniform mesh of size $h$.
We define the mesh P\'{e}clet numbers $$\beta=\sigma h/2, \quad \gamma=\tau h/2, \quad \delta=\mu h/2. $$
We use the skew-symmetric part of this matrix (that is, given $A$, form $\frac{A-A^T}{2}$) for our skew-symmetric experiments.

In our tests, we have found that equilibration has not been particularly effective. We speculate that this might have to do with a property related to block diagonal dominance that these matrices have for certain values of the convective coefficients. Specifically, the norm of the tridiagonal part of the matrix is significantly larger than the norm of the remaining part. Equilibration tends to adversely affect this property by scaling down entries near the diagonal, and as a result the performance of an iterative solver often degrades. We thus do not apply equilibration in our skew-symmetric solver.

In Table \ref{tab:skew1} we manipulate the drop tolerance for ILDL, to obtain a fill nearly equal to that of ILUTP. For the latter we fix the drop tolerance at 0.001.  This is done for the purpose of comparing the performance of the iterative solvers, when the memory requirements of ILUTP and ILDL are similar. Prior to preconditioning, we apply AMD as a fill-reducing reordering. We apply preconditioned GMRES(100) to solve the linear system, until either a residual of $10^{-6}$ is reached, or until 1000 iterations are used. If the linear system fails to converge after 1000 iterations, we mark it as NC. We see that the iteration counts are significantly better for ILDL, especially when rook pivoting is used. Note that our ILDL still consumes only about 2/3 of the memory of ILUTP, due to the fact that the floating point entries of only half of the matrix are stored.

\renewcommand{\arraystretch}{1.3}
\begin{table}[ht]

\tbl{Comparison of {\sc Matlab}'s ILUTP and SYM-ILDL for a skew-symmetric matrix arising from a model convection-diffusion equation \label{tab:skew1}}{
\begin{tabular}{lllllcc}
\hline
n & ${\rm nnz}(A)$ & method & drop tol &  fill & GMRES(20) & time (s) \\
\hline
\multirow{3}{*}{$20^3=8000$} & \multirow{3}{*}{$45600$} & ILDL-rook & $10^{-4}$ & 7.008 & 6 & 0.130+0.041 \\
& & ILDL-partial & $5\cdot10^{-4}$ & 6.861 & 6 & 0.138+0.041 \\
& & ILUTP & $10^{-3}$ & 7.758 & 8 & 0.406+0.038 \\ \hline
\multirow{3}{*}{$30^3=27000$} & \multirow{3}{*}{$156600$} & ILDL-rook & $2\cdot10^{-4}$ & 10.973 & 8 & 0.936+0.246 \\
& & ILDL-partial & $3\cdot10^{-4}$ & 11.235 & 10 & 1.162+0.331 \\
& & ILUTP & $10^{-3}$ & 11.758 & 13 & 4.475+0.307 \\ \hline
\multirow{3}{*}{$40^3=64000$} & \multirow{3}{*}{$374400$} & ILDL-rook & $9\cdot10^{-5}$ & 15.205 & 9 & 3.820+0.855 \\
& & ILDL-partial & $3\cdot10^{-4}$ & 15.686 & 18 & 4.971+1.582 \\
& & ILUTP & $10^{-3}$ & 15.654 & 19 & 26.63+1.40 \\ \hline
\multirow{3}{*}{$50^3=125000$} & \multirow{3}{*}{$735000$} & ILDL-rook & $2\cdot10^{-5}$ & 21.560 & 6 & 15.39+1.76 \\
& & ILDL-partial & $2\cdot10^{-4}$ & 22.028 & 62 & 21.11+17.95 \\
& & ILUTP & $10^{-3}$ & 22.691 & 58 & 151.14+11.60 \\ \hline
\multirow{3}{*}{$60^3=216000$} & \multirow{3}{*}{$1274400$} & ILDL-rook & $2\cdot10^{-5}$ & 22.595 & 9 & 34.82+4.02 \\
& & ILDL-partial & $4\cdot10^{-4}$ & 22.899 & NC & 36.17+NC \\
& & ILUTP & $10^{-3}$ & 23.483 & NC & 356.60+NC\\ \hline
\multirow{3}{*}{$70^3=343000$} & \multirow{3}{*}{$2028600$} & ILDL-rook & $5\cdot10^{-6}$ & 32.963 & 5 & 106.81+3.25 \\
& & ILDL-partial & $4\cdot10^{-4}$ & 36.959 & NC & 156.52+NC\\

& & ILUTP & $10^{-3}$ & 33.861 & NC & 876.33+NC  \\
\hline
\end{tabular}}

\begin{tabnote}
The parameter used were $\beta=20, \gamma=2, \delta=1$. The {\sc Matlab} ILUTP used a drop tolerance of 0.001. `NC' stands for `no convergence'.
\end{tabnote}
\end{table}

%
%

\begin{figure}[ht]

\begin{center}
	\includegraphics[width=0.45\textwidth]{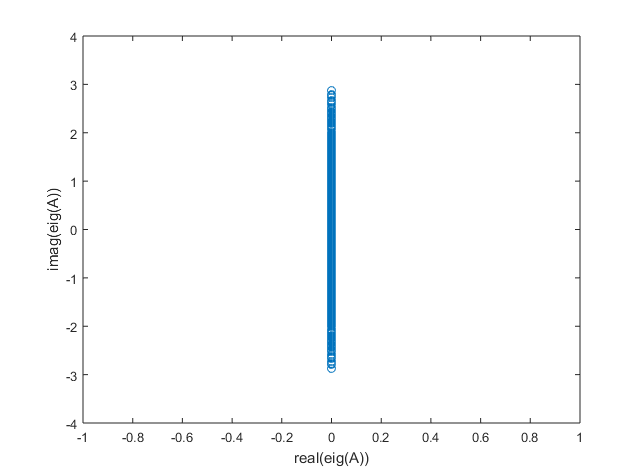}
\includegraphics[width=0.45\textwidth]{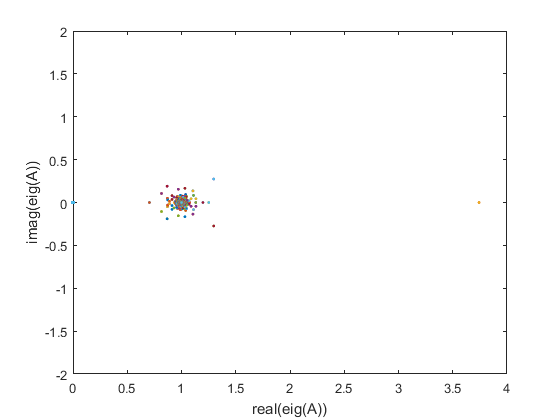}
\end{center}

\caption{Eigenvalues of an unpreconditioned (left) and preconditioned (right) skew-symmetric $1000 \times 1000$ matrix $A$ arising from a convection-diffusion model problem. \label{fig:eigskew}}
\end{figure}

In Figure \ref{fig:eigskew} we show the (complex) eigenvalues of the preconditioned matrix $(L D L^T)^{-1} A$, where $A$ is the skew-symmetric part of \ref{eq:convdiff} with convective coefficients $(\beta,\gamma,\delta)=(0.4,0.5,0.6)$, and $L D L^T$ is the preconditioner generated by running SYM-ILDL with a drop tolerance of $10^{-3}$ and a fill-in parameter of $20$.

For the purpose of comparison, we also show the unpreconditioned eigenvalues. As seen in the figure, most of the preconditioned eigenvalues are very strongly clustered around 1, which indicates that a preconditioned iterative solver is expected to rapidly converge. The unpreconditioned eigenvalues are pure imaginary, and follow the formula $$2\left(\beta \cos\left(j \pi h \right) + \gamma \cos\left(k \pi h \right) + \delta \cos\left(\ell \pi h \right)\right)\imath $$ where $1\leq j,k,\ell \leq 1/h$.

\section{Obtaining and Contributing to SYM-ILDL}
SYM-ILDL is open source, and documentation can be found at {\tt \url{http://www.cs.ubc.ca/~greif/code/sym-ildl.html}}. We essentially allow free use of our software with no restrictions. To this end, SYM-ILDL uses the MIT Software License.

We welcome any contributions to our code. Details on the contribution process can be found with the link above. Certainly, more code optimization is possible, such as parallelization; such tasks remain as items for future work.

\section*{Acknowledgments} We would like to thank Dominique Orban and Jennifer Scott for their careful reading of an earlier version of this manuscript, and Yousef Saad for providing us with the code from \cite{lsc2005}. We would also like to thank the referees for their helpful and constructive comments.

\bibliographystyle{ACM-Reference-Format-Journals}
\bibliography{ghl2016_arxiv}

\begin{appendix}

\begin{table}[ht]

\tbl{Factorization timings and iterative solver iterations for test matrices \label{tab:all-results}}{
\begin{tabular}{lllllll}
\hline
matrix  & $n$  & ${\rm nnz}(A)$  & fill  &  time (s) & type & iterations\tabularnewline
\hline
\multirow{5}{*} {aug3dcqp}  & \multirow{5}{*} {$35543$}  & \multirow{5}{*} {$128115$}  & 1.9& 0.051+0.148 & ILDL(B+AMD) & 24 \tabularnewline
& & &  3.3     & 0.063+0.442 & ILDL(MC64+MC64R) & 55\\
& & &  2.1     & 0.068+0.261 & ILDL(MC64+AMD) & 33\\
& & &  3.2     & 0.063+0.223 & ILDL(B+MC64R) & 33\\
& & & & & &\\
& & &  7.3 & 2.655+0.198 & ILUTP(B+AMD) & 6\\
& & &  21.2 & 11.674+0.890 & ILUTP(MC64+MC64R) & 14\\
& & &  36.0 & 11.513+0.397 & ILUTP(MC64+AMD) & 6\\
& & &  7.4 & 1.753+0.215 & ILUTP(B+MC64R) & 7\\
\hline
\multirow{5}{*} {bloweya}  & \multirow{5}{*} {$30004$}  & \multirow{5}{*} {$150009$}  & 0.9 & 0.030+0.081 & ILDL(B+AMD) & 18 \tabularnewline
& & &  1.0     & 0.071+0.014 & ILDL(MC64+MC64R) & 3\\
& & &  1.0     & 0.023+0.019 & ILDL(MC64+AMD) & 5\\
& & &  0.9     & 0.152+0.126 & ILDL(B+MC64R) & 18\\
& & & & & &\\
& & & 2.8 & 38.817+0.101 & ILUTP(B+AMD) & 4\\
& & & 6.1 & 2.726+0.109 & ILUTP(MC64+MC64R) & 4\\
& & & 2.9 & 39.537+0.104 & ILUTP(MC64+AMD) & 4\\
& & & 3.2 & 7.858+0.100 & ILUTP(B+MC64R) & 3\\
\hline
\multirow{5}{*} {bratu3d}  & \multirow{5}{*} {$27792$}  & \multirow{5}{*} {$173796$}  & 3.8 & 0.358+0.155 & ILDL(B+AMD) & 23\tabularnewline
& & &  3.6     & 0.155+0.124 & ILDL(MC64+MC64R) & 24\\
& & &  3.6     & 0.231+0.272 & ILDL(MC64+AMD) & 36\\
& & &  3.8     & 0.245+0.105 & ILDL(B+MC64R) & 18\\
& & & & & &\\
& & & 8.6 & 22.237+0.214 & ILUTP(B+AMD) & 9\\
& & & 10.3 & 13.114+0.962 & ILUTP(MC64+MC64R) & 18\\
& & & 9.8 & 32.717+0.500 & ILUTP(MC64+AMD) & 10\\
& & & 8.1 & 8.480+0.540 & ILUTP(B+MC64R) & 11\\
\hline
\multirow{5}{*} {tuma1}  & \multirow{5}{*} {$22967$}  & \multirow{5}{*} {$87760$}  & 2.9& 0.044+0.201 & ILDL(B+AMD) & 50\tabularnewline
& & &  3.0     & 0.051+0.132 & ILDL(MC64+MC64R) & 35\\
& & &  3.0     & 0.077+0.299 & ILDL(MC64+AMD) & 54\\
& & &  3.0     & 0.046+0.220 & ILDL(B+MC64R) & 59\\
& & & & & &\\
& & &       7.8 & 2.686+0.582 & ILUTP(B+AMD) & 14\\
& & &       40.0 & 19.476+0.495 & ILUTP(MC64+MC64R) & 8\\
& & &       20.7 & 7.268+0.242 & ILUTP(MC64+AMD) & 6\\
& & &       17.7 & 7.750+51.991 & ILUTP(B+MC64R) & NC\\
\hline
\multirow{5}{*} {tuma2}  & \multirow{5}{*} {$12992$}  & \multirow{5}{*} {$49365$}  &2.8 & 0.023+0.084 & ILDL(B+AMD) & 41\tabularnewline
& & &  3.0     & 0.029+0.087 & ILDL(MC64+MC64R) & 28\\
& & &  3.0     & 0.045+0.104 & ILDL(MC64+AMD) & 34\\
& & &  3.0     & 0.041+0.218 & ILDL(B+MC64R) & 55\\
& & & & & &\\
& & &       6.9 & 0.720+0.226 & ILUTP(B+AMD) & 13\\
& & &       33.8 & 4.140+0.192 & ILUTP(MC64+MC64R) & 7\\
& & &       19.0 & 1.936+0.106 & ILUTP(MC64+AMD) & 5\\
& & &       15.5 & 2.082+12.341 & ILUTP(B+MC64R) & 697\\
\hline

\end{tabular}}

\end{table}

\begin{table}[ht]
\ContinuedFloat

\captionsetup{labelformat=empty}
\tbl{}{
\begin{tabular}{lllllll}

\multirow{5}{*} {boyd1}  & \multirow{5}{*} {$93279$}  & \multirow{5}{*} {$1211231$}  & 1.0 & 0.155+0.077 & ILDL(B+AMD) & 3 \tabularnewline
& & &  0.6     & 0.102+0.505 & ILDL(MC64+MC64R) & 42\\
& & &  1.0     & 0.123+0.088 & ILDL(MC64+AMD) & 6\\
& & &  0.6     & 0.144+0.437 & ILDL(B+MC64R) & 36\\
& & & & & &\\
& & &       0.8 & 0.219+1.021 & ILUTP(B+AMD) & 10\\
& & &       0.8 & 0.257+0.875 & ILUTP(MC64+MC64R) & 12\\
& & &       0.8 & 0.233+1.656 & ILUTP(MC64+AMD) & 14\\
& & &       0.8 & 0.188+0.481 & ILUTP(B+MC64R) & 10\\
\hline

\multirow{5}{*} {brainpc2}  & \multirow{5}{*} {$27607$}  & \multirow{5}{*} {$179395$}  & 1.0 & 0.878+0.094 & ILDL(B+AMD) & 31 \tabularnewline
& & &  1.8     & 0.315+0.100 & ILDL(MC64+MC64R) & 31\\
& & &  1.5     & 1.661+0.085 & ILDL(MC64+AMD) & 28\\
& & &  1.8     & 0.481+0.983 & ILDL(B+MC64R) & 214\\
& & & & & &\\
& & &       0.6 & 0.541+38.711 & ILUTP(B+AMD) & NC\\
& & &       961.5 & 373.210+1210.140 & ILUTP(MC64+MC64R) & NC\\
& & &       88.7 & 15.434+180.070 & ILUTP(MC64+AMD) & NC\\
& & &       0.6 & 0.925+38.263 & ILUTP(B+MC64R) & NC\\
\hline

\multirow{5}{*} {mario001}  & \multirow{5}{*} {$38434$}  & \multirow{5}{*} {$204912$}  & 3.7 & 0.163+0.541 & ILDL(B+AMD) & 54 \tabularnewline
& & &  3.6     & 0.234+0.629 & ILDL(MC64+MC64R) & 55\\
& & &  3.6     & 0.213+0.603 & ILDL(MC64+AMD) & 54\\
& & &  3.7     & 0.129+0.557 & ILDL(B+MC64R) & 52\\
& & & & & &\\
& & &       8.0 & 2.474+0.542 & ILUTP(B+AMD) & 8\\
& & &       9.3 & 26.39+0.612 & ILUTP(MC64+MC64R) & 8\\
& & &       9.0 & 2.552+0.555 & ILUTP(MC64R+AMD) & 8\\
& & &       8.6 & 21.73+0.325 & ILUTP(B+MC64) & 9\\
\hline
\multirow{5}{*} {qpband}  & \multirow{5}{*} {$20000$}  & \multirow{5}{*} {$45000$}  & 1.1 & 0.008+0.004 & ILDL(B+AMD) & 1 \tabularnewline
& & &  1.1     & 0.007+0.004 & ILDL(MC64+MC64R) & 1\\
& & &  1.8     & 0.014+0.004 & ILDL(MC64+AMD) & 1\\
& & &  1.1     & 0.016+0.004 & ILDL(B+MC64R) & 1\\
& & & & & &\\
& & &       1.1 & 0.008+0.026 & ILUTP(B+AMD) & 1\\
& & &       1.1 & 0.008+0.021 & ILUTP(MC64+MC64R) & 1\\
& & &       1.2 & 0.011+0.028 & ILUTP(MC64+AMD) & 1\\
& & &       1.1 & 0.010+0.013 & ILUTP(B+MC64R) & 1\\
\hline

\end{tabular}}

\end{table}

\begin{table}[ht]
\ContinuedFloat

\captionsetup{labelformat=empty}
\tbl{}{
\begin{tabular}{lllllll}

\multirow{5}{*} {nlpkkt80}  & \multirow{5}{*} {$1062400$}  & \multirow{5}{*} {$28192672$}  & 9.5 & 113+1308 & ILDL(B+AMD) & 998*\tabularnewline
& & &   14.5  & 176+1580 & ILDL(MC64+MC64R) & 854*\\
& & &   12.3  & 153+53 & ILDL(MC64+AMD) & 34\\
& & &   10.6  & 121+NC & ILDL(B+MC64R) & NC\\
& & & & & &\\
& & &   4.1   & 6803 + 2502& ILUTP(B+AMD) & NC\\
& & &   -  & - & ILUTP(MC64+MC64R) & -\\
& & &   -  & - & ILUTP(MC64+AMD) & -\\
& & &   -  & - & ILUTP(B+MC64) & -\\
\hline
\multirow{5}{*} {nlpkkt120}  & \multirow{5}{*} {$3542400$}  & \multirow{5}{*} {$95117792$}  & 9.8 & 401+NC & ILDL(B+AMD) & NC\tabularnewline
& & &   14.5 & 533+NC & ILDL(MC64+MC64R) & NC\\
& & &   12.4    & 525+334 & ILDL(MC64+AMD) & 58\\
& & &   10.9 & 460+NC & ILDL(B+MC64R) & NC\\
& & & & & &\\
& & &   -  & - & ILUTP(B+AMD) & -\\
& & &   -  & - & ILUTP(MC64+MC64R) & -\\
& & &   -  & - & ILUTP(MC64+AMD) & -\\
& & &   -  & - & ILUTP(B+MC64) & -\\
\hline
\end{tabular}}

\begin{tabnote}
The experiments were run with {\tt fill\_factor} = 2.0 for the smaller matrices and {\tt fill\_factor} = 4.0 for matrices larger than one million in dimension. The tolerance was
{\tt drop\_tol} = $10^{-4}$, and we used rook pivoting to maintain stability. The iteration was terminated when the norm of the relative residual went below $10^{-6}$. For iteration counts labelled with a *, MINRES was used (as SQMR failed to converge). Iteration counts labelled with NC indicates non-convergence for both MINRES and SQMR. Times labelled with `-' took over 10 hours to run, and were terminated before completion.
\end{tabnote}

\end{table}

\setcounter{table}{6}

\begin{table}[h]

The table below uses {\tt HSL\_MC80} on matrices from Table 1 and 3 in this paper. For MC80, we chose AMD as the fill-reducing reordering after the matching stage. Only matrices from these two tables were used as all other matrices in our tests were well-scaled and block-diagonally dominant to begin with (such as the Helmholtz problem of Table 2).

\tbl{Results with HSL\_MC80 for matrices in Table 1 and 3 \label{tab:mc80}}{
\begin{tabular}{lllllll}
\hline
matrix  & $n$  & ${\rm nnz}(A)$  & fill  &  time (s) & iterations\tabularnewline
\hline
{aug3dcqp}  & {$35543$}  & {$128115$}  & 2.0 & 0.051+0.188 & 28 \tabularnewline

{bloweya}  & {$30004$}  & {$150009$}  & 0.9 & 0.036+0.023 & 4 \tabularnewline

{bratu3d}  & {$27792$}  & {$173796$}  & 2.9 & 0.118+0.106 & 26\tabularnewline

{tuma1}  & {$22967$}  & {$87760$}  & 3.0& 0.063+0.227 & 44\tabularnewline

{tuma2}  & {$12992$}  & {$49365$}  &2.9 & 0.033+0.094 & 35\tabularnewline

{boyd1}  & {$93279$}  & {$1211231$}  & 1.0 & 0.120+0.062 & 4 \tabularnewline

{brainpc2}  & {$27607$}  & {$179395$}  & 1.5 & 0.086+0.119 & 26 \tabularnewline

{mario001}  & {$38434$}  & {$204912$}  & 3.6 & 0.110+0.501 & 59 \tabularnewline

{qpband}  & {$20000$}  & {$45000$}  & 1.1 & 0.015+0.004 & 1 \tabularnewline
\hline

{nlpkkt80}  & {$1062400$}  & {$28192672$}  & 7.1 & 133+86 & 49\tabularnewline

{nlpkkt120}  & {$3542400$}  & {$95117792$}  & - & x+x & -\tabularnewline
\hline

c-55  & 32780  & 403450  & 2.95 & 0.28+0.15 & 15\tabularnewline
c-59  & 41282  & 480536  & 2.99 & 0.36+0.20 & 15\tabularnewline
c-63  & 44234  & 434704  & 2.92 & 0.29+0.21 & 15\tabularnewline
c-68  & 64810  & 565996  & 2.31 & 0.31+0.17 & 9\tabularnewline
c-69  & 67458  & 623914  & 2.65 & 0.35+0.18 & 9\tabularnewline
c-70  & 68924  & 658986  & 2.67 & 0.40+0.24 & 11\tabularnewline
c-71  & 76638  & 859520  & 3.00 & 0.74+0.32 & 12\tabularnewline
c-72  & 84064  & 707546  & 2.69 & 0.40+0.21 & 9\tabularnewline
c-big & 345241 & 2340859 & 2.54 & 1.2+0.93 & 8\tabularnewline
\end{tabular}}

\begin{tabnote}
Matrices in the first section (delimited by horizontal lines) were run with {\tt fill\_factor} = 2.0 and {\tt drop\_tol} = $10^{-4}$. Matrices in the second section  were run with {\tt fill\_factor} = 4.0 and {\tt drop\_tol} = $10^{-4}$. Matrices in the third section were run with {\tt fill\_factor} = 12.0 and {\tt drop\_tol} = $0.003$. Rook pivoting was used to maintain stability. The iterative solvers used for the first two sections was SQMR, and GMRES was used for the third section. These settings maintain consistency with Tables 1 and 3 of Section 5. The iteration was terminated when the norm of the relative residual went below $10^{-6}$. Iteration counts labelled with NC indicates non-convergence.
\end{tabnote}

\end{table}

\begin{table}[ht]

\tbl{GMRES comparisons between SYM-ILDL and AMD with MC64 equilibration\label{tab:scott} }{
\begin{tabular}{c|cc|ccc|ccc}
\hline
Matrix name & $n$ & ${\rm nnz}(A)$ & $\text{fill}_\text{MI30}$ & MI30 iters & time (s) & $\text{fill}_\text{SYM-ILDL}$ & SYM-ILDL iters & time (s) \tabularnewline
\hline
c-55  & 32780  & 403450  & 3.45 & 49 & 1.25+0.94 & 3.85 & 12 & 0.49+0.15\tabularnewline
c-59  & 41282  & 480536  & 3.62 & 70 & 1.59+1.84 & 3.70 & 13 & 0.59+0.27\tabularnewline
c-63  & 44234  & 434704  & 4.10 & 51 & 1.53+1.23 & 4.12 & 13 & 0.48+0.25\tabularnewline
c-68  & 64810  & 565996  & 4.12 & 37 & 1.87+1.12 & 4.00 & 9 & 0.69+0.26\tabularnewline
c-69  & 67458  & 623914  & 4.33 & 43 & 4.07+1.47 & 3.93 & 11 & 0.64+0.34\tabularnewline
c-70  & 68924  & 658986  & 4.26 & 38 & 3.77+1.30 & 3.46 & 13 & 0.58+0.42\tabularnewline
c-71  & 76638  & 859520  & 3.58 & 61 & 3.93+2.71 & 4.09 & 10 & 1.13+0.40\tabularnewline
c-72  & 84064  & 707546  & 4.18 & 54 & 3.05+2.40 & 5.33 & 14 & 1.15+0.59\tabularnewline
c-big & 345241 & 2340859 & 4.82 & 67 & 23.4+25.3 & 2.92 & 11 & 1.89+1.62\tabularnewline
\hline
\end{tabular}}

\begin{tabnote}
For each test case, we report the time it takes to compute the preconditioner, as well as the GMRES time and the number of GMRES iterations. The time is reported as $x+y$, where $x$ is the preconditioning time and $y$ is the GMRES time. GMRES was terminated when the relative residual decreased below $10^{-6}$.
\end{tabnote}

\end{table}

\end{appendix}

\end{document}